# My Experience in Physical Layer Communications

Xiang-Gen Xia

University of Delaware

I feel that I have been very lucky since I have experienced the most dynamic 30 years on electronics in the past. I think that the most visible change in our daily life over the past 30 years is communications. From computer modems, to internet, and to smart phones, people now feel much less lonely or bored since they are always connected. In this article, I would like to share with you on my own experience working on communications in the past decades. To do so, I would like to start from my graduate student life in 1983.

**When I was a Graduate Student**

My undergraduate major was pure mathematics and the major I wanted to work on for my master (MS) degree was Information Theory - Signal Processing in the Mathematics Department, Nankai University that was one of the best universities in China in the 1980s. The main reason for me was that the major course in the graduate entry exams was Real Analysis that was what I liked the most in my college mathematics studies and Nankai was the best university among those with Real Analysis as the major exam course. Another reason was that signal processing was towards applications, although I did not know what signal processing or information theory was exactly.

After I was luckily admitted to Nankai University as a MS student in the Fall of 1983, I learned that my supervisor, Prof. Guoding Hu, on the admission paper work was one of the pioneers in information theory in China. He studied information theory with Dobrushin, a world leading expert in Russia, in the 50s. However, my major was signal processing that was, in fact, different from the information theory in Nankai University at that time. Although that was the case, I still listened the Information Theory course but I did not understand it well. I did not know what communications was about either. During the five years in Nankai University, I studied signal processing but my main interests were still in mathematics and I did not fully understand what signal processing was really about, since I did not have any electrical engineering background or knowledge. I did some math research on some mathematics problems in signal processing.

In the Fall of 1990, I came to University of Southern California (USC) to study for my Ph.D. degree in electrical engineering (systems). At that time, USC was one of the major towns in communications in the world and had world leading experts in most areas in communications, such as Reed, Welch, Golomb, and Kumar in error control coding, my Ph.D. supervisor, Dr. Zhen Zhang, in information theory, Scholtz, Lindsey, and Weber et al. in communications systems, Dr. Victor Li in networks, and Gagliardi and Willner in optical communications. Among them, six are members of National Academy of Engineering (NAE) in USA and three received the Shannon Award.

After I came to USC, I started to learn some basic courses in electrical engineering – systems, such as, information theory, digital communications, digital signal processing, digital image processing, and error correction coding etc. After the studies of these courses, I started to understand what communications is about. In the meantime, I started to do research with my supervisor on error correction coding mainly involving combinatorics. During the early 1990s, communications was not hot anymore and one hot topic was wavelets and its applications in signal and image processing. Since my MS major was in signal processing in Nankai University, I had always paid my attention on, at least, some of the mathematics problems in signal processing. So, in addition to doing research on error correction coding, I worked on wavelets that, in fact, became my Ph.D. thesis topic.

I remember that before I graduated at USC, my officemate-classmate-friend and myself came across to the new book by Biglieri et al. on trellis coded modulation (TCM) that was the first book on TCM. We xerox-copied the book but had a hard time to understand its major concept called coding gain. During the early 1990s, although communications was cool down, algebraic geometry error correction coding (AG codes) had attracted a lot attention in error correction coding area in the world, where algebraic geometry, such as Riemann-Roch Theorem, plays the key role. This was a hot area in information theory community. As we all know that AG codes became cool down after turbo coding was invented and attracted much attention worldwide in 1993 and from 1995, respectively. Also note that AT&T was split again and was split to three companies in 1995.

Another popular area in communications and signal processing in the first half of 1990s was image compression, because of the standards JPEG and MPEG. I remember that every university had multiple departments and many research groups working on this topic using various tools, such as wavelets, vector quantization (VQ), fractal theory, and neural networks. It is in source coding area in information theory and text lossless compression was popular as well. One of the most famous techniques is the discrete cosine transform (DCT) image compression that was started in USC in the 1970s using one of the early digital images, Lena, from the Playboy magazine. It became one of the key technologies in JPEG and MPEG2. Later, wavelet tree image compression became one of the key technologies in JPEG 2000. Note that all these standards are used in our current image and video files these days, which one may not realize.

**When I Worked in Industry**

I became truly interested in communications in and since 1995 after I joined Hughes Research Lab. (HRL). HRL belonged to Hughes Aircraft Company that has made the best communications stationary satellites in the world, and also provided the first commercial digital television broadcasting, i.e., DirecTV in 1994. In HRL, I worked on some projects on code division multiple access (CDMA) code designs and MPEG4. I started to well understand all the concepts I mentioned earlier and more concepts and technologies in communications, for example, intersymbol interference (ISI) and coding gain in TCM. Since in the past I was working on wavelets and multirate filterbanks, I naturally applied the filterbank theory to ISI channels as precoding without knowing the ISI channel, which turns out to be able to get rid of the spectral nulls in theory by adding a little bit redundancy at the transmitter [1].

When I followed MPEG4 in 1995 and 1996, there were three news reported in big media, such as New York Times (NYT) and Wall Street Journal (WSJ), on new image compression methods. One was from Texas A &M group using wavelets, one was from Yale Group using VQ (I think), and one was from Gatech Group using fractals. All these three groups had their own companies. At least two of these groups claimed that they could compress 100 times for images of size 256 by 256 with 256 grey levels, which was a big claim, since I did image compression by myself and it was even hard for me to have compression ratio 16. There were two times of reports on NYT (or WSJ and I do not remember exactly which one it was now) for the Texas Group but none of them had any details, which made a lot people curious about. In the beginning I did not believe them. However, after many repeated claims/sayings/rumors, somehow I tended to believe them as well. Of course, later none of the three was found real, unfortunately.

I worked in HRL for exactly one and half years during 1995 and 1996. HRL is located in Malibu and seating on the middle of a mountain. My office was facing Malibu beach with Malibu village of Hollywood actors and actresses. I was fascinated in the beginning but quickly got tired of seeing the Pacific Ocean waves every day.

**When I Worked in Academia**

I continued working on filterbank precoding for ISI channels after I joined University of Delaware (UDel) in 1996, and found with Hui Liu that after adding the minimum redundancy at the transmitter without knowing ISI channel, one is able to blindly equalize an ISI channel at the receiver. We call these precoders ambiguity resistant precoders [2,3,4].

I also treated filterbank precoders as complex field coding, where the input signals are complex modulated information symbols and the arithmetic operations between the input and the linear coding are complex number arithmetic operations (note that the conventional error control coding uses binary or finite field arithmetic operations between an input signal and its coding). I call it modulated coding. I found that such a complex field linear coding does not have any coding gain for the AWGN channel, i.e., the ideal channel, which may not be surprising although. However, I found that it may have coding gain for an ISI channel, even compared to the uncoded AWGN channel [6]. In other words, for the AWGN channel, one does not need to do any linear complex field coding because it does not gain anything, but it may be helpful for an ISI channel.

In the meantime, I got involved in orthogonal frequency division multiplexing (OFDM) systems, since one can treat OFDM as a linear complex field coder. By combining some of the filterbank theory, I proposed vector OFDM for single antenna systems [5]. Vector OFDM is in the middle of OFDM (when the vector size is 1, i.e., the scalar case) and single carrier frequency domain equalizer (SC-FDE) (when the vector size is at least the ISI channel length), both of which have been used in LTE/4G as downlink and uplink, respectively. At the receiver side, OFDM is able to convert an ISI channel to multiple ISI-free sub-channels, while for SC-FDE, the number of symbols interfering each other is the ISI channel length. For vector OFDM, the number of symbols interfering each other is the vector size if the vector size is less than the ISI channel length, and the vector size can be chosen according to whatever the need is. So, in terms of ISI, vector OFDM is a bridge between OFDM and SC-FDE. The maximum-likelihood (ML) demodulation complexity of vector OFDM is also in the middle of those of OFDM and SC-FDE.

All these precoding results are summarized in my book [6], which covers three cases. One case is when neither transmitter nor receiver knows ISI channel, which corresponds to ambiguity resistant precoders. One case is when only receiver knows ISI channel, which corresponds to vector OFDM. The third case is when both transmitter and receiver know ISI channel, where one is able to find the optimal modulated code at the transmitter to have the highest coding gain for a given ISI channel.

After I finished the work on vector OFDM in 1999, I, with my graduate students, started to work on space-time coding for MIMO systems. Luckily, we obtained numerous results on orthogonal space-time codes, quasi-orthogonal space-time codes, lattice space-time codes, space-time codes with simplified decoding, unitary space-time codes, and space-time codes for relay channels etc. Some of these results, such as those on complex orthogonal designs and unitary matrices, are new in mathematics [7].

**Summary in Physical Layer Modulations**

When we look back the past 70 years, we may find that one of the major tasks in communications physical layer has been always on how to deal with ISI. This was particularly important for computer modem design over telephone line in the 1990s. It combines two of the most important technologies, one of which is at the transmitter side, i.e., the bandwidth efficient coding, namely TCM, and the other is at the receiver side, i.e., the decision feedback equalizer (DFE). Computer modem design was the most important business in communications in the 1990s and produced multiple standards and products on the market. To further improve the data speed, wireline high speed modems using OFDM (or multi tone) were developed. This is wireline communications.

In wireless communications, it is the same, i.e., the major task in physical layer is still on how to deal with ISI. Cellular and WiFi are the two most common wireless communications systems. Let us use cellular systems as examples. For cellular systems, it has migrated from 1G to 5G now. Since 1G was analog, let us start with 2G. I think that a wireless communications system mainly depends on its bandwidth. The wider bandwidth a channel is, the more multipaths the channel has or the longer the ISI channel is. For 2G, its bandwidth is about 1.3 MHz, which can be thought of as a narrowband system. In this case, the ISI length is short, i.e., the ISI is not severe, both TDMA and CDMA systems work well as we have seen from Europe GSM to North America IS-95 standards, respectively.

However, when bandwidth is higher, such as the bandwidth in 3G, around 10MHz, the number of multipaths, or the number of ISI symbols, becomes significant. In this case, the time domain equalization cannot be fast enough to adapt to a wireless fading channel in a TDMA system. In contrast, in a CDMA system, when the CDMA codeword length is not too short, the mainlobe of a CDMA codeword is still significantly higher than the sum of all the sidelobes produced from the ISI of the channel. Thus, the chip level RAKE receiver works well to deal with the chip level

ISI. This is exactly the reason why all 3G systems are CDMA based, such as CDMA2000 and WCDMA.

When the bandwidth further increases, such as in 4G/LTE, it is 20MHz, CDMA may not work well. In this case, the number of multipaths or the number of ISI symbols, becomes large. On the other hand, to have a high data rate, a CDMA codeword cannot be too long. Thus, when the number of multipaths is large, the mainlobe of a CDMA codeword may not be higher than the sum of all the sidelobes produced from the ISI of the channel. Therefore, the chip level RAKE receiver may not work well. This is the reason why in 4G/LTE, CDMA is not used while OFDM is used, where OFDM converts an ISI channel to multiple ISI free channels. Since the channel bandwidth for a WiFi system is also 20MHz, similar to the 4G cellular system, WiFi uses OFDM as well. For more details about the above argument, see [8].

**Past Milestones in Physical Layer Communications**

When we think about the physical layer digital communications, I think that the milestones are as follows.

Of course, the most important foundation is Shannon information theory, i.e., both channel coding and source coding theorems, which includes the upper and the lower bounds. The upper bound is the channel capacity that is the transmission data rate upper bound for a given channel. The lower bound is the entropy bound for lossless data compression and rate distortion theory for lossy data compression, which are the minimum data rate to have for a given source and a given distortion in the lossy data compression case. In Chinese, it is 如来在世.

The source coding includes both lossless and lossy data compressions. The milestones:  for lossless compression, they are Huffman coding and Lempel-Ziv-Welch algorithm; and for lossy compression, they are (vector) quantization, code-excited linear prediction (CELP) coding for speech and audio compressions, DCT and discrete wavelet transform (DWT) algorithms for image and video compressions. In Chinese, it is 精益求精.

For the channel coding, I think that the milestones are: Reed-Solomon and BCH codes, Viterbi decoding algorithms, trellis coded modulation (TCM), and the iterative decoding in turbo codes and LDPC codes. In Chinese, it is 海底捞针.

For modulations, the milestones are: CDMA, OFDM, and Alamouti coding for multiple antenna systems. In Chinese, it is 排兵布阵.

For communication systems and/or signal processing, the milestones include phase locked loop synchronization, matched filtering, and decision feedback equalizer (DFE). In Chinese, it is 画龙点睛.

**Conclusion**

Research is luxury. Among massive research results and papers, only a very few can survive or be kept as a few golden pins dropped and found in the ocean. These few milestones in communications and the milestones in electronics hardware development, such as chips, have changed our daily life communications in the past 30 years dramatically. In the meantime, as a side benefit, the massive research efforts have trained and are training a lot young students who are invaluable to a society.

**PS:** After my reading of papers and listening of seminars on communications these days, I intended to write a complain about the current physical layer communications. However, it has ended up with a memory in the past.

**References**

[1] X.-G. Xia, "New precoding for intersymbol interference cancellation using nonmaximally decimated multirate filterbanks with ideal FIR equalizers," *IEEE Trans. on Signal Processing*, vol. 45, no. 10, pp. 2431-2441, Oct. 1997.

[2] H. Liu and X.-G. Xia, "Precoding techniques for undersampled multi-receiver communication systems," *IEEE Trans. on Signal Processing*, vol. 48, pp, 1853-1863, Jul. 2000.

[3] X.-G. Xia and H. Liu, "Polynomial ambiguity resistant precoders: theory and applications in ISI/multipath cancellation," *Circuits, Systems, and Signal Processing*, vol.19, no.2, pp.71-98, 2000.

[4] X.-G. Xia, W. Su, and H. Liu, "Filterbank precoders for blind equalization: Polynomial ambiguity resistant precoders (PARP)," *IEEE Trans. on Circuits and Systems I*, vol. 48, no. 2, pp. 193-309, Feb. 2001.

[5] X.-G. Xia, "Precoded and vector OFDM robust to channel spectral nulls and with reduced cyclic prefix length in single transmit antenna systems," *IEEE Trans. on Communications*, vol. 49, no. 8, pp. 1363-1374, Aug. 2001.

[6] X.-G. Xia, *Modulated Coding for Intersymbol Interference Channels*, New York, Marcel Dekker, Oct. 2000.

[7] X.-G. Xia, "Mathematics and electrical engineering," Nankai Univ., Tianjin, China, June 28, 2018. [Online]. Available: https://www.eecis.udel.edu/~xxia/Math_EE.pdf.

[8] X.-G. Xia, "Topics on modulations for 5G and beyond," University of Delaware, Newark, Delaware, USA, 2017. [Online] available: https://www.eecis.udel.edu/~xxia/modulations_beyond_5G.pdf.